\documentclass[pre,final,twocolumn,superscriptaddress,showpacs,footinbib,floatfix]
{revtex4}
\usepackage{amsmath}
\usepackage{epsfig}
\newcommand{\ket}[1]{\left| #1 \right\rangle}
\newcommand{\bra}[1]{\left\langle #1 \right|}

\newcommand{\sinc}{\mathrm{sinc}}

\begin{document}
\bibliographystyle{revtex}

\title{Efficiency of Free Energy Calculations of Spin Lattices by Spectral
Quantum Algorithms}
\author{Cyrus P. Master}
\email{cpmaster@stanford.edu}
\affiliation{Quantum Entanglement Project, ICORP, JST,
Stanford University, Stanford, CA 94305-4085}
\author{Fumiko Yamaguchi}
\affiliation{Quantum Entanglement Project, ICORP, JST,
Stanford University, Stanford, CA 94305-4085}
\author{Yoshihisa Yamamoto}
\affiliation{Quantum Entanglement Project, ICORP, JST,
Stanford University, Stanford, CA 94305-4085}
\affiliation{NTT Basic Research Laboratories,
3-1 Morinosato-Wakamiya, Atsugi, Kanagawa 243-0198, Japan}
    \date{\today}
\begin{abstract}
Quantum algorithms are well-suited to calculate estimates of the
energy spectra for spin lattice systems.  These algorithms are
based on the efficient calculation of discrete Fourier components
of the density of states.  The efficiency of these algorithms in
calculating the free energy per spin of general spin lattices to
bounded error is examined. We find that the number of Fourier
components required to bound the error in the free energy due to
the broadening of the density of states scales polynomially with the
number of spins in the lattice.  However, the precision with which
the Fourier components must be calculated is found to be an
exponential function of the system size.
\end{abstract}
\maketitle

\section{Introduction}

Spin lattice models are useful for the study of magnetic ordering
in real materials.  The dynamics of these models are specified by
a Hamiltonian $\hat{\mathcal{H}}$ involving spin operators for
each of the $n$ lattice sites.  Of particular interest is the
behavior of thermodynamic functions -- such as the magnetization,
specific heat capacity, and magnetic susceptibility -- across
phase transitions. These functions are encapsulated in the
dependence of the Helmholtz free energy per spin, $F$, on system
parameters such as the temperature or applied magnetic field;
partial derivatives of $F$ yield the thermodynamic functions.
Thus, calculation of $F$ over a wide parameter space suffices for
the determination of the finite temperature behavior of the spin
lattice model.

Calculation of the free energy for a general spin lattice by
conventional means is difficult.  A naive approach is to enumerate
the eigenenergies $\{E_m\}$ of $\hat{\mathcal{H}}$, since
\begin{equation}
F = -n^{-1}k_B\theta\ln Z = -n^{-1}k_B\theta\ln
\left(\sum_m e^{-\beta E_m}\right),
\label{Eq:I.1}
\end{equation}
where $k_B\theta = \beta^{-1}$ is the thermal energy, and $Z$ is
the partition function.  However, as the number of eigenstates
grows exponentially with the number of spins in the lattice, the
time required to perform the calculation is prohibitively large. A
variety of quantum Monte Carlo methods exist to calculate the free
energy, including thermodynamic integration
\cite{Frenkel1,deKoning1}, histogram methods
\cite{Ferrenberg1,Alves1} and cumulant expansion
\cite{Rickman1,Phillpot1} techniques.  However, the ``sign
problem'' (see, for example, Ref. \onlinecite{Landau1}) prevents
application of these methods to arbitrary lattice Hamiltonians.


An alternate approach is available if one can efficiently generate
an estimate of the density of states $\rho(E)$.  As Eq.
(\ref{Eq:I.1}) may be written in the form
\begin{equation}
F = -n^{-1}k_B\theta\ln\left(\int_{-\infty}^\infty\rho(E)e^{-\beta
E}dE\right),
\end{equation}
an approximation for the density of states $\rho(E)$ directly
translates into an estimate $\tilde{F}$ for the free energy per
spin.

Algorithms for quantum computers have been proposed to determine
information about the spectra of Hermitian operators
\cite{Shor2,Kitaev1,Cleve1,Abrams1,DeRaedt1,Somma1}.  We focus on
algorithms \cite{DeRaedt1,Somma1} that efficiently generate
estimates of individual Fourier components $f_\ell$ of $\rho(E)$;
they will be reviewed in detail in Section II.  $N$ iterations of
the algorithms yield $N$ Fourier components, from
which an estimate of the density of states can be calculated.

An important issue that has not been addressed is the efficiency
of these algorithms for calculating thermodynamic functions as a
function of $n$.  For the calculation to be deemed efficient, it
must be shown that the computation time -- and, thus, the number
of Fourier components -- required to calculate an estimate $\tilde{F}$
to bounded error scales polynomially with $n$.  The bounded error criterion
we adopt is
\begin{equation}
\mathrm{Prob}\left(|\tilde{F}-F| <
\gamma k_B\theta\right) > 1-\epsilon, \label{Eq:I.3}
\end{equation}
where $\gamma$ and $\epsilon$ are small constants.  Thus, the
absolute error in the estimated free energy per spin must be
smaller than a fraction of the thermal energy with probability
arbitrarily close to one.

We examine the primary sources of error involved in the
calculation of $F$ to determine the efficiency of the
spectral algorithms.  First, as only a finite number of Fourier
components $f_\ell$ of the density of states are calculated, the
estimated density of states is broadened relative to the actual
function. This \emph{deterministic} source of error (\emph{i.e.},
it is unchanged if the calculation is repeated) is reduced by
increasing the number of components $N$, and thus the computation
time.  Second, there is an inherent \emph{stochastic} source of
error reflected in deviations in the estimated $f_\ell$ from the
actual values.  This error could arise from imprecise
implementation of logic gates or noise in the measurement process.

In this paper, we will show that if all of the $f_\ell$ are known
exactly, the bound in Eq. (\ref{Eq:I.3}) may be met by a number of
Fourier components that scales polynomially with $n$.  Thus, the
error due to the broadening of the density of states does not
prevent efficient estimation of the free energy per spin. However,
the free energy becomes increasingly sensitive to random errors in
each of the $f_\ell$ as the number of spins is increased.  We will
show that the precision of the output of the quantum algorithm
must improve exponentially with $n$ in order to sustain the
condition in Eq. (\ref{Eq:I.3}). Thus, it is questionable as to
whether spectral algorithms can be applied to the calculation of
thermodynamic functions.

The paper is organized as follows: Section II reviews the quantum
algorithms used to generate the components $f_\ell$, and discusses
assumptions and expected properties of the spin Hamiltonian.
Section III describes the calculation of $\tilde{F}$ from the
Fourier components, and discusses the influence of sampling and
window functions.  In Section IV, we analyze the deterministic
error due to broadening and determine the number of samples
required to meet Eq. (\ref{Eq:I.3}).  In Section V, we analyze the
impact of random deviations in the components $f_\ell$ on the
estimated free energy.

\section{Quantum Algorithms}
In this section, we review quantum algorithms for the calculation
of the Fourier transform of the density of states.  We
describe a simple algorithm applicable only to Hamiltonians that
are diagonal in the computational basis, and then discuss a more
general algorithm \cite{Knill1} applicable to ensemble quantum computers.

In regards to notation, we use the standard model for quantum
computation, assuming our $p$ qubits to be two-level systems with
logical states $\ket{0}_j$ and $\ket{1}_j$,
$j\in\{0,1,\ldots,p-1\}$, corresponding to eigenstates of the
$\hat{\sigma}_z^{(j)}$ Pauli spin operator with eigenvalues $\pm
1$.  The computational basis states for the quantum computer are
denoted as $\ket{x}=\ket{x_1}_1\ket{x_2}_2\ldots\ket{x_p}_p$,
where $\{x_j\}$ are the binary digits of the integer $x$. It is
assumed that the quantum computer is capable of implementing a
universal set of elementary single-qubit and two-qubit gates.  The
evolution time of these gates is an implementation-dependent
constant, such that the overall computation time is reflected by
the number of gates used in the algorithm.

We will restrict our discussion to lattices of spin-$1/2$ particles, as it
is straightforward to map the eigenstates of $\hat{\sigma}_z^{(j)}$ in
the spin system to the logical $\ket{0}_j$ and $\ket{1}_j$ states
of qubit $j$ of the quantum computer.  It should be noted that this restriction
does not preclude the treatment of lattices of particles with spins larger
than $1/2$.  Generalized Jordan-Wigner transformations exist
\cite{Batista1,Jordan1} to represent the dynamics of such lattices by
a collection of spin-1/2 particles \emph{via} intermediate fermionization.

Prior to the discussion of individual algorithms, we state three
assumptions regarding the nature of the spin lattice Hamiltonian.
First, we assume that the energy bandwidth $\Delta E$ -- the
energy difference between the ground state and the highest excited
state -- is bounded by a polynomial function of $n$.  This
assumption is likely to be valid for models of interest.  As an
example, consider a lattice of particles interacting by an
nearest-neighbor XXZ interaction:
\begin{equation}
\hat{\mathcal{H}} = \sum_{\left<i,j\right>}\left[J_x\left(\hat{\sigma}_x^{(i)}
\hat{\sigma}_x^{(j)}+\hat{\sigma}_y^{(i)}
\hat{\sigma}_y^{(j)}\right)+J_z\hat{\sigma}_z^{(i)}
\hat{\sigma}_z^{(j)}\right]. \label{Eq:II.1}
\end{equation}
The expectation value of the summand in Eq. (\ref{Eq:II.1}) must
lie between $-(2|J_x|+|J_z|)$ and $(2|J_x|+|J_z|)$.  The number of
terms in the summation is equal to $n/2$ times the coordination
number for the lattice, implying that the energy bandwidth is
bounded by a function linear in $n$.  In general, for any
Hamiltonian involving only pairwise interactions \footnote{This
assumption is more restrictive if a pair-wise Hamiltonian for a
lattice of particles with spin larger than $1/2$ is transformed to
an equivalent spin-1/2 lattice, as the transformed Hamiltonian
will not necessarily consist of only pair-wise interactions.}
between spins (of $n$-independent interaction energy), it is
evident that the energy bandwidth is $O(n^2)$.

Second, we assume that the time-evolution operator
$\hat{U}(t)\equiv\exp(-i\hat{\mathcal{H}}t)$ can be implemented as a
sequence of elementary single-qubit and two-qubit gates, where the
number of gates is a polynomial function of $n$.  In cases where
the Hamiltonian consists of commuting pair-wise interactions
(\emph{e.g.}, the Ising model), this decomposition is trivial.
Otherwise, one may use a Trotter-Suzuki expansion of non-commuting
terms \cite{Suzuki1,Lloyd1} to implement $\hat{U}(t)$ to
arbitrarily small error.

Finally, we assume that the energy scale is shifted such that the
eigenenergies fall between $E=0$ and $E=\Delta E$.  This last
assumption is made for mathematical convenience, and does not
affect the results of our analysis.

The following algorithms are based on the fact that the Fourier transform
of the density of states $\rho(E)$ is equal to the trace of the
time evolution operator:
\begin{align}
f(t) &\equiv
\int_{-\infty}^{\infty}\rho(E)e^{-iEt}dE =
\mathrm{Tr}\left(e^{-i\hat{\mathcal{H}}t}\right).
\end{align}
As $|f(t)|\leq 2^n$, it is convenient to define a function
\begin{equation}
g(t) \equiv \frac{1}{2^n}f(t) = \frac{1}{2^n}\mathrm{Tr}
\left(e^{-i\hat{\mathcal{H}}t}\right),
\label{eq:def g(t)}
\end{equation}
such that $|g(t)|\leq 1$.  The algorithms described in this section
calculate samples of $g(t)$ at discrete times $t_\ell$.

Before discussing the general case, it is illuminating to examine
a simple algorithm restricted to spin lattices for which $\hat{\mathcal{H}}$
is diagonal in the computational basis.  As an example, one might consider
a nearest-neighbor Ising model in a longitudinal magnetic field:
\begin{equation}
\hat{\mathcal{H}} =
J_z\sum_{\{i,j\}}\left(1-\hat{\sigma}_z^{(i)}\hat{\sigma}_z^{(j)}\right)
+h\sum_i\left(1-\hat{\sigma}_z^{(i)}\right). \label{eq: Ising
model}
\end{equation}
The gates shown in Fig. 1 for an $n$-qubit computer can be used to
calculate the magnitude of $g({t_\ell})\equiv g_\ell$.  The
quantum computer is initialized to the $\ket{0}$ state.  The gate
$W$ corresponds to a sequence of Walsh-Hadamard gates $W_j$ for
each qubit $j$:
\begin{equation}
W_j: \left\{
\begin{array}{ll}
\ket{0}_j \rightarrow \frac{1}{\sqrt{2}}
\left(\ket{0}_j+\ket{1}_j\right) \\
\ket{1}_j \rightarrow \frac{1}{\sqrt{2}}
\left(\ket{0}_j-\ket{1}_j\right)
\end{array}.
\right.
\end{equation}
The gate $U({t_\ell})$ corresponds to the time evolution operator.
\begin{figure}
\begin{center}
\epsfxsize = 3in \epsfbox{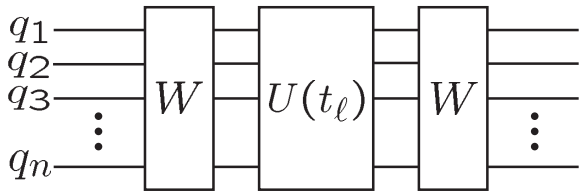}
\caption{Logic diagram of an elementary algorithm to estimate $|g_\ell|$.}
\end{center}
\label{fig: 1}
\end{figure}

As a final step, a projective measurement is performed in the computational
basis. It is straightforward to show that the probability of observing all
qubits in the logical $\ket{0}$ state is equal to $|g_\ell|^2$:
\begin{align}
\ket{0}&\xrightarrow{W}\frac{1}{\sqrt{2^n}}\sum_{m=0}^{2^n-1}\ket{m}
\xrightarrow{U({t_\ell})}\frac{1}{\sqrt{2^n}}\sum_{m=0}^{2^n-1}
e^{-iE_m {t_\ell}}\ket{m} \nonumber \\ &\xrightarrow{W}
g_\ell\ket{0}+\text{orthogonal components}.
\end{align}
By assumption, the computational basis states $\ket{m}$ are
eigenstates of the Hamiltonian, and the time-evolution operator
appends a phase proportional to the eigenvalue $E_m$ to each term.
An unbiased estimator for $|g_\ell|$ can be derived by
performing $R$ repetitions of the algorithm $R$, and counting the number
of times all qubits are found in the logical $\ket{0}$ state.

The magnitude of $g_\ell$ is insufficient to reconstruct the density of
states.  By adding an ancilla qubit $a$, as shown in Fig. 2, one may extract
estimates of both the real and imaginary parts of $g_\ell$.  The $X
\equiv\exp\left(i\pi\hat{\sigma}_x^{(a)}/4\right)$ and $Y \equiv
\exp\left(i\pi\hat{\sigma}_y^{(a)}/4\right)$ gates correspond to
$\pi/2$ rotations of the ancilla qubit.  If the $X$ gate is
used, the probabilities of observing the
$\ket{\phi_0}\equiv\ket{0}_a\ket{0}_{q_1}\ldots\ket{0}_{q_n}$ or
$\ket{\phi_1}\equiv\ket{1}_a\ket{0}_{q_1}\ldots\ket{0}_{q_n}$ states
are
\begin{align}
p_{X0} = \left|\frac{1+ig_\ell}{2}\right|^2, \
p_{X1} = \left|\frac{1-ig_\ell}{2}\right|^2,
\end{align}
respectively.  The $Y$ gate leads to probabilities
\begin{align}
p_{Y0} = \left|\frac{1+g_\ell}{2}\right|^2, \
p_{Y1} = \left|\frac{1-g_\ell}{2}\right|^2.
\end{align}
By executing $R$ iterations of the algorithm with the $X$ gate and $R$
iterations with the $Y$ gate, one can derive estimators $\tilde{p}$
for the probabilities.  Unbiased estimates of the real and imaginary
parts of $g_\ell$ are
\begin{align}
\mathrm{Re}\left(\tilde{g}_\ell\right) &=
\left(\tilde{p}_{Y0}-\tilde{p}_{Y1}\right), \\
\mathrm{Im}\left(\tilde{g}_\ell\right) &=
\left(\tilde{p}_{X1}-\tilde{p}_{X0}\right).
\end{align}
We use the tilde to distinguish estimates of the Fourier components
obtained from the quantum algorithm from the exact values.
\begin{figure}
\begin{center}
\epsfxsize = 3in \epsfbox{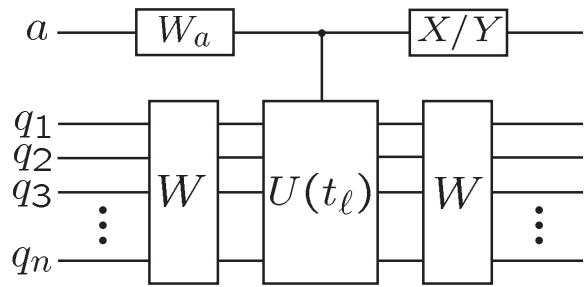}
\caption{Logic diagram of an algorithm to estimate the real and imaginary
parts of $g_\ell$ for diagonal $\hat{\mathcal{H}}$.}
\end{center}
\label{fig: 2}
\end{figure}

The algorithm described above depends on our ability
to construct an equally-weighted coherent superposition of eigenstates of
$\hat{\mathcal{H}}$; hence, the restriction to Hamiltonians that are
diagonal in the computational basis.  It can be generalized
\cite{Knill1} by considering an algorithm involving an ensemble of quantum
computers, such that the ancilla qubit is still initialized to
the $\ket{0}_a$ state, but the remaining $n$ qubits are in a completely
random mixed state.  The initial density matrix for the system
is
\begin{equation}
\hat{\rho}_i = \frac{1}{2^{n+1}}\left(\hat{I}^{(a)}+{\hat{\sigma}_z^{(a)}}
\right)\hat{I}^{(q_1)}\hat{I}^{(q_2)}\ldots\hat{I}^{(q_n)},
\end{equation}
where $\hat{I}^{(\ell)}$ is the identity operator for qubit $\ell$.  The
operator $\hat{I}^{(q_1)}\hat{I}^{(q_2)}\ldots\hat{I}^{(q_n)}$
is equal to the resolution of the identity
$\sum_{\ell}\ket{\psi_\ell}\bra{\psi_\ell}$, where $\{\ket{\psi_\ell}\}$ is
an orthogonal set of states in the subspace spanned by qubits $q_1$ through
$q_n$.  One could use as $\{\ket{\psi_\ell}\}$ the eigenstates of the
Hamiltonian $\hat{\mathcal{H}}$.  We do not need to explicitly solve for the
eigenstates of $\hat{\mathcal{H}}$; the initial density matrix can be thought of
as an incoherent mixture of eigenstates for any Hamiltonian $\hat{\mathcal{H}}$.

If the coherent superposition created by the Walsh-Hadamard gates
is replaced by such an incoherent mixture, then an algorithm
nearly identical to the one shown above works for any choice of
$\hat{\mathcal{H}}$, as shown in Fig. 3.  The final measurement is
the expected value of $\hat{\sigma}_z^{(a)}$ averaged over the
ensemble.
\begin{figure}
\label{fig: 3}
\begin{center}
\epsfxsize = 3in \epsfbox{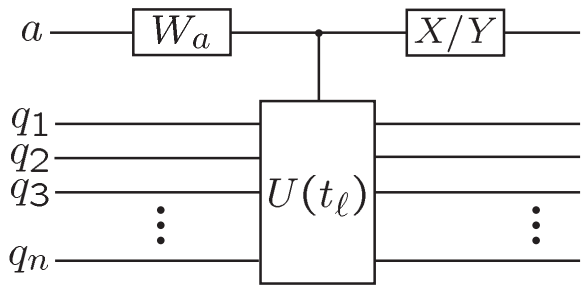}
\caption{Logic diagram of an ensemble algorithm to determine $g_\ell$ for
arbitrary $\hat{\mathcal{H}}$.}
\end{center}
\end{figure}

If the $X$ gate is used for the ancilla qubit, the expected value of
$\hat{\sigma}_z^{(a)}$ is
\begin{equation}
\left<\hat{\sigma}_z^{(a)}\right> =
\mathrm{Im}\left(g_\ell\right),
\end{equation}
whereas the $Y$ gate leads to
\begin{equation}
\left<\hat{\sigma}_z^{(a)}\right> =
\mathrm{Re}\left(g_\ell\right).
\end{equation}
Thus, the real and imaginary components of the estimator
$\tilde{g}_\ell$ can be calculated from two iterations of the
ensemble quantum algorithm.

The ensemble algorithm is attractive for two reasons.  First, it
is applicable to any spin-1/2 lattice Hamiltonian
$\hat{\mathcal{H}}$, provided that the time-evolution operator can
be decomposed into a sufficiently small number of elementary
gates.  Second, initial state preparation lends itself to ensemble
quantum computation proposals involving spin resonance, where the
Z\'{e}eman splitting between qubit spin states is small compared
to the thermal energy. In equilibrium, the initial density matrix
of the system is well-approximated by the identity operator.  The
single pseudopure state qubit can be created from two thermal
spins \cite{Gershenfeld1}.

\section{Free Energy Estimation}
In this section, we discuss how an estimate $\tilde{F}$ of the
free energy is generated from the Fourier components of the
density of states, and we examine the effects of discretization on the
estimated density of states.

A conceptual overview of the free energy calculation including
post-processing is shown in Fig. 4.  $N$ samples of $f(t)$ are
estimated \emph{via} the quantum algorithm \footnote{As this section
exploits the Fourier relationship between $f(t)$ and the density of states,
it is more convenient notation-wise to commence with the former than $g(t)$.},
and are weighted by a windowing function $b_\Theta(t)$, described below.  Fourier
transformation yields an estimate $\tilde{\rho}(E)$ for the
density of states. The density of states may be integrated to compute
the partition function, and, thus, the free energy.

As iterations of the quantum algorithm yield discrete samples of
$f(t)$, the reconstructed estimate $\tilde{\rho}(E)$ is distorted
relative to the exact function $\rho(E)$.  This distortion
translates into error in $\tilde{F}$.  It is convenient to
view this error in the context of windowing and sampling of the
exact Fourier transform $f(t)$ of the density of states.
Truncation of $f(t)$ to a window of width $T_0$ centered about
$t=0$ (\emph{i.e.}, multiplication of $f(t)$ by a windowing function
$b_1(t)$ that is constant for $|t| \leq T_0/2$ and zero elsewhere)
leads to a convolution of the density of states by a broadening
function $b_1(E) \equiv \alpha_1\sinc\left(\frac{\pi E}{\Delta
e}\right) =\alpha_1\left[\sin\left(\frac{\pi E}{\Delta
e}\right)\right] /\left(\frac{\pi E}{\Delta e}\right)$, where the
energy resolution $\Delta e$ is given by
\begin{equation}
\Delta e = \frac{2\pi}{T_0}. \label{E3:energy resolution window
size}
\end{equation}
The window is scaled such that the broadening function is
normalized to unit area; \emph{i.e.,} $\alpha_1 = 1/\Delta e$.
Increasing the window size $T_0$ reduces the width of the broadening
function, and thus the error in the estimate of $\tilde{F}$.
\begin{figure}
\begin{center}
\epsfxsize = 3in \epsfbox{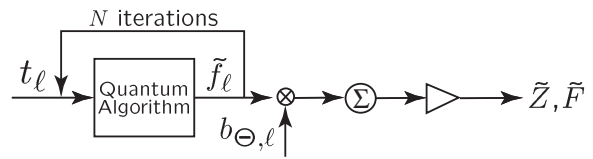}
\caption{Block diagram of the calculation of $\tilde{F}$.}
\end{center}
\label{fig: 4}
\end{figure}

The effect of sampling on the estimated density of states can be
determined by multiplying $f(t)b_1(t)$ by an impulse train $s(t)$
of spacing $\Delta t$:
\begin{equation}
s(t) = \Delta t\sum_{\ell=-\infty}^\infty
\delta(t-\ell\Delta t).
\label{Eq:sdef}
\end{equation}
Sampling leads to periodic replication of the broadened $\rho(E)$.  The
resultant density of states is given by the Fourier transform of
$f(t)b_1(t)s(t)$:
\begin{equation}
\rho'(E) \equiv \rho(E) \ast b_1(E) \ast \sum_{k=-\infty}^\infty
\delta\left(E+\frac{2\pi k}{\Delta t}\right).
\label{eq: estimated DOS as convolution}
\end{equation}
To avoid aliasing in the estimated density of states, the Nyquist
sampling condition requires that
\begin{equation}
\Delta t \leq \frac{2\pi}{\Delta E}. \label{E3: energy bandwidth
sampling time}
\end{equation}
The spacing between samples of $f(t)$ is determined solely by
the estimate of the energy bandwidth, $\Delta E$.  We
assume that sampling is performed at the Nyquist rate, and the
equality holds in Eq. (\ref{E3: energy bandwidth sampling time}).

As the number of samples is equal to the ratio of the windowing
function width $T_0$ to the sampling time, one could determine the
minimum value of $T_0$ required to satisfy Eq. (\ref{Eq:I.3}) as a
function of $n$.  However, the rectangular windowing function
leads to poor results.  The envelope of the associated broadening
function $b_1(E)$ falls off weakly as $1/E$; the oscillating side
lobes are amplified at low energies by the Boltzmann
factor in the calculation of the free energy.  The window width
required to mitigate the resultant error scales poorly with $n$.
In contrast to using wider rectangular windows, one may
adopt more elaborate window shapes whose corresponding
broadening functions exhibit envelopes that are more sharply
peaked.  We consider the functions $b_\Theta(t)$ formed by the
successive convolution of $\Theta$ rectangular windows, each of
width $T_0$. $\Theta$ is referred to as the order of the windowing
function. For $\Theta=2$, the window is triangular and of width $2
T_0$.  With increasing order, the window approaches a Gaussian
shape, and is of width $\Theta T_0$. The resulting broadening
function is then
\begin{equation}
b_\Theta(E) = \alpha_\Theta \left[\sinc\left(\frac{\pi E}{\Delta
e}\right)\right]^\Theta, \label{E3: defn of b_Theta}
\end{equation}
which exhibits a $1/E^\Theta$ envelope.  The value of
$\alpha_\Theta$ is determined, as the area under
$b_\Theta(E)$ is one. In practice, a given window shape is
constructed by obtaining samples $\tilde{f}_\ell$ within the window
width $\Theta T_0$ centered at $t=0$, and weighting each sample by
$b_{\Theta,\ell}\equiv b_\Theta({t_\ell})$.

As the envelope of the side lobes of $b_\Theta(E)$ falls off
exponentially with $\Theta$, windowing functions of large order
significantly reduce the error in the calculated free energy.
However, the tradeoff is a wider window, leading to more Fourier
components, and thus more iterations of the quantum algorithm:
\begin{equation}
N=\frac{\Theta T_0}{\Delta t}.
\label{eq: N samples}
\end{equation}
Therefore, the question of how $N$ scales with the number of spins, $n$,
translates into the determination of the minimum values of $\Theta$ and $T_0$
required to satisfy Eq. (\ref{Eq:I.3}).

An estimate $\tilde{Z}$ for the partition function can be
calculated directly from the estimated Fourier components without
intermediate calculation of the density of states.  We denote
quantities obtained from the quantum algorithm with a tilde, in
contrast to their exact values.  First, note that the Fourier
transform of $\tilde{f}(t)b_\Theta(t)s(t)$ may be evaluated explicitly
via Eq. (\ref{Eq:sdef}) to give an estimate of the broadened, periodically
replicated density of states $\rho'(E)$ in terms of the components
$\tilde{f}_\ell$:
\begin{align}
\tilde{\rho}'(E) &= \frac{1}{2\pi}\int_{-\infty}^\infty \tilde{f}(t)b_\Theta(t)
\left[\Delta t \sum_{\ell=-\infty}^\infty
\delta(t-\ell\Delta t)\right]e^{iEt}dt \nonumber \\
&= \frac{\Delta t}{2\pi}\sum_{\{t_\ell\}} \tilde{f}_\ell b_{\Theta,\ell}
e^{iEt_\ell},
\label{rho from f samples}
\end{align}
where we have defined $t_\ell\equiv\ell\Delta t$, and the sum is performed
over all $t_\ell$ within the window described by $b_{\Theta,\ell}$.
Integrating Eq. (\ref{rho from f samples}) over the
energy bandwidth and using Eqs. (\ref{eq:def g(t)}) and
(\ref{E3: energy bandwidth sampling time}), one finds
\begin{widetext}
\begin{align}
\tilde{Z} &= \int_0^{\Delta E}\tilde{\rho}'(E)e^{-\beta E} dE
= \frac{2^n\Delta t}{2\pi\beta}\left(1-e^{-\beta\Delta E}\right)
\left\{b_{\Theta,0}+2\sum_{\ell>0}^{N/2}b_{\Theta,\ell}
\left[\frac{1}{1+(t_\ell/\beta)^2}\mathrm{Re}(\tilde{g}_\ell)
-\frac{t_\ell/\beta}{1+(t_\ell/\beta)^2}\mathrm{Im}(\tilde{g}_\ell)\right]
\right\}.
\label{partition function from f samples}
\end{align}
\end{widetext}
Note that an estimate $\tilde{F}$ of the free energy may be obtained from the
logarithm of Eq. (\ref{partition function from f samples}).

In addition to describing how an estimate of the free energy per spin is
calculated from the Fourier components, Eq.
(\ref{partition function from f samples}) will serve as a starting point to
determine the probabilistic error in $\tilde{F}$ due to imprecise values
of $\tilde{g}_\ell$.

\section{Error Analysis: Broadening}

In this section, we determine an upper bound on the number of
samples $N$ of $g(t)$ required to calculate the free energy to the tolerance
prescribed by Eq. (\ref{Eq:I.3}).  At this point, we consider the individual
samples of $g_\ell$ to be known exactly, and only consider the error in
$\tilde{F}$ due to the finite number of Fourier components --
\emph{i.e.}, due to broadening of the density of states.  With this
restriction, we can show that $N$ is a polynomial function of the number of
spins $n$.

As it is more convenient to work with the partition function than the free
energy, we use a more stringent bound based upon the relative error
in the calculated partition function $\tilde{Z}$.  As
\begin{equation}
\left|\tilde{F}-F\right| < \gamma k_B \theta \Longleftrightarrow
e^{-\gamma n}-1 < \frac{\tilde{Z}-Z}{Z} < e^{\gamma n}-1,
\end{equation}
it is sufficient to demand that
\begin{equation}
\mathrm{Prob}\left(r\equiv\left|\frac{\tilde{Z}-Z}{Z}\right| < \xi\right)
> 1-\epsilon, \label{eq: Z error criterion}
\end{equation}
where $\xi\equiv 1-\exp(-\gamma n)$.  Note that $\xi<1$, and in the limit
$\gamma n \ll 1$, $\xi\rightarrow\gamma n$.

By Eqs. (\ref{E3:energy resolution window size}), (\ref{E3: energy bandwidth
sampling time}) and (\ref{eq: N samples}), if the Nyquist sampling condition
is satisfied, then
\begin{equation}
N = \frac{\Theta\Delta E}{\Delta e}.
\label{eq: N}
\end{equation}
It has been asserted that $\Delta E$ is a polynomial function of $n$.
In the remainder of this section, we examine the dependence of
$\Theta$ and $\Delta e$ on $n$ such that Eq. (\ref{eq: Z error criterion})
is satisfied.  We require a pair of intermediate results:

\boxed{\emph{Lemma 1:}} If $b_\Theta(E)$ (as defined in Eq.
(\ref{E3: defn of b_Theta})) is subject to the normalization
condition $1 = \int_{-\infty}^\infty b_\Theta(E)dE$, then
\begin{equation}
\alpha_\Theta < \frac{c\pi}{\Delta e}\sqrt{\frac{\Theta}{6\pi}},
\label{E3: Lemma 1}
\end{equation}
where $c \approx 2.0367$.

\boxed{\emph{Lemma 2:}}
\begin{equation}
A_{\text{side}} \equiv 1-\int_{-\Delta e}^{\Delta e} b_\Theta(E)dE <
\frac{c}{\pi^{\Theta-3}} \sqrt{\frac{\Theta}{6\pi}}, \label{E3:
Lemma 2}
\end{equation}
where $\Theta$ is an even integer.

Lemma 1 places an upper bound on $\alpha_\Theta$ such
that $b_\Theta(E)$ is normalized.  Lemma 2 states an upper bound on
the area of $b_\Theta(E)$ that is outside of the
interval $[-\Delta e,\Delta e]$ (\emph{i.e.},
outside the main lobe of the broadening function) that decreases
exponentially with $\Theta$. Both are proved in the Appendix.  Note that
the proof of Lemma 2 applies to the case where $\Theta$ is even.

One can relate the relative error $r$ in the
calculated partition function to the parameters $\Theta$ and
$\Delta e$ \emph{via} Lemma 2.  As the exact density of states $\rho(E)$ may
be expressed as a sum of delta functions for each eigenenergy $E_m$,
Eqs. (\ref{eq: estimated DOS as convolution})-(\ref{E3: energy bandwidth
sampling time}) at the Nyquist condition yield
\begin{align}
\tilde{Z} &= \int_0^{\Delta E}\tilde{\rho}(E)
e^{-\beta E}dE \nonumber \\
&= \sum_m \sum_{k=-\infty}^\infty \int_0^{\Delta E}
b_\Theta(E-E_m+k\Delta E)e^{-\beta E}dE \nonumber \\
&= \sum_m \left[\sum_{k=-\infty}^\infty \int_{k\Delta E}^{(k+1)\Delta E}
b_\Theta(E-E_m)e^{-\beta(E-k\Delta E)}dE\right] \nonumber \\
&\equiv \sum_m \tilde{Z}_m.
\end{align}
The change of variables allows one to view $\tilde{Z}_m$ as an integral of
the broadening function, centered at $E_m$, and weighted by
periodically-replicated segments of an exponential function.  $\tilde{Z}$
is found by summing over all eigenenergies.

The maximum relative error $r$ in the partition function is bounded by the
largest contribution $r_m \equiv \max_m \left|\frac{\tilde{Z}_m-Z_m}{Z_m}
\right|$ from any single eigenenergy, where $Z_m \equiv e^{-\beta E_m}.$
Define $\gamma_m = \tilde{Z}_m/Z_m$. Then,
\begin{align}
r &= \left|\frac{\tilde{Z}-Z}{Z}\right| = \frac{\left|\sum_m
\left(\tilde{Z}_m-Z_m\right)\right|}{\sum_m Z_m} =
\frac{\left|\sum_m(\gamma_m-1)Z_m\right|}{\sum_m Z_m}
\nonumber \\
&\leq \max_m \left|\gamma_m - 1\right| = \max_m
\left|\frac{\tilde{Z}_m-Z_m}{Z_m}\right|=r_m.
\end{align}

This argument shows that one may consider a simplified system with just one
eigenstate at an energy ${E_m}$ somewhere in the energy bandwidth. An upper
bound on the error $r$ for this simplified system for any ${E_m}$ suffices to
bounded the error for an arbitrary energy spectrum over the same
bandwidth.

Lower and upper bounds on $\tilde{Z}_m$ ($\tilde{Z}_{m,\text{min}}$
and $\tilde{Z}_{m,\text{max}}$, respectively) are now derived to bound
$r_m$, since
\begin{equation}
r_m < \max\left(\left|\frac{\tilde{Z}_{m,\text{min}}-Z_m}{Z_m}\right|,
\left|\frac{\tilde{Z}_{m,\text{max}}-Z_m}{Z_m}\right|\right). \label{E3:
ri}
\end{equation}
In the main lobe, the minimum value of the Boltzmann factor
is $e^{-\beta({E_m}+\Delta e)}$.  Outside
of the main lobe, the minimum value is $e^{-\beta\Delta E}$.
Thus,
\begin{align}
\tilde{Z}_m &= \sum_{k=-\infty}^\infty \int_{k\Delta E}^{(k+1)\Delta E}
b_\Theta(E-E_m)e^{-\beta(E-k\Delta E)}dE \nonumber \\
&\geq (1-A_{\text{side}})e^{-\beta({E_m}+\Delta e)}+A_{\text{side}}
e^{-\beta\Delta E} \equiv \tilde{Z}_{m,\text{min}}. \label{E3: Zimin}
\end{align}
Similarly, as the maximum value of the Boltzmann factor is
$e^{-\beta({E_m}-\Delta e)}$ inside the main lobe and one
outside,
\begin{align}
\tilde{Z}_m  \leq (1-A_{\text{side}})e^{-\beta({E_m}-\Delta
e)}+A_{\text{side}} \equiv \tilde{Z}_{m,\text{max}}. \label{E3: Zimax}
\end{align}
Substituting Eqs. (\ref{E3: Zimin}) and (\ref{E3: Zimax}) into Eq.
(\ref{E3: ri}), we see that
\begin{align}
r_m < \max[&1-(1-A_{\text{side}})e^{-\beta\Delta e}
-A_{\text{side}}e^{-\beta(\Delta E-{E_m})}, \nonumber \\
&(1-A_{\text{side}})e^{\beta\Delta e}+
A_{\text{side}}e^{\beta{E_m}}-1]. \label{E3: ri2}
\end{align}
It is difficult to invert Eq. (\ref{E3: ri2}) explicitly to find
optimal conditions on $A_{\text{side}}(\Theta)$ and $\Delta e$ that ensure
that $r_m < \xi$. However, one can show that the following
conditions are sufficient:
\begin{align}
\beta\Delta e &= \ln(1+\xi/2), \label{E3: Delta e condition}\\
A_{\text{side}} &< \frac{\xi}{2}e^{-\beta\Delta E}. \label{E3: Aside
condition}
\end{align}
As proof of their sufficiency, note that
\begin{align}
&1-(1-A_{\text{side}})e^{-\beta\Delta e}-A_{\text{side}}e^{-\beta(\Delta
E-{E_m})} \nonumber \\
&< \frac{\xi}{2}+\frac{\xi}{2}e^{-\beta\Delta E}\left(1-\frac{\xi}{2}\right)
\nonumber \\
&< \xi,
\end{align}
and,
\begin{align}
&(1-A_{\text{side}})e^{\beta\Delta e}+A_{\text{side}}e^{\beta{E_m}}-1
\nonumber \\
&< \frac{\xi}{2}+\frac{\xi}{2}e^{-\beta(\Delta E-E_m)}
\nonumber\\
&< \xi.
\end{align}
Therefore, the conditions in Eqs. (\ref{E3: Delta e
condition}) and (\ref{E3: Aside condition}) guarantee that $r < r_m <
\xi$, as desired.

Using Lemma 2, one can manipulate Eqs. (\ref{E3: Delta e
condition}) and (\ref{E3: Aside condition}) to show that $N$ scales
polynomially with $n$.
\begin{align}
\Delta e &= \frac{\ln(1+\xi/2)}{\beta}, \label{E3: final condition 1}\\
\Theta -
\frac{\ln\Theta}{2\ln\pi}&> \frac{\beta\Delta E}{\ln \pi}+
\frac{\ln(1/\xi)}{\ln\pi}+\kappa, \label{E3: final condition 2a}
\end{align}
where $\kappa = 5/2+\frac{\ln(2c/\sqrt{6})}{\ln\pi}
\approx 2.9443$ .  As $\ln\Theta < \Theta$, a sufficient condition to
satisfy Eq. (\ref{E3: final condition 2a}) is
\begin{equation}
\Theta/2 = \lceil\mu\beta\Delta E + \mu\ln(1/\xi) + \kappa'\rceil,
\label{eq: final condition 2b}
\end{equation}
where $\mu \equiv 1/(2\ln\pi-1)$ and $\kappa' \equiv \mu\kappa\ln\pi$.

In summary, the error bound on the partition function is satisfied if the
energy resolution scales linearly with temperature, and if $\Theta$
scales linearly with $\beta\Delta E$.

As a final step, we substitute the conditions in Eqs.
(\ref{E3: final condition 1}) and (\ref{eq: final condition 2b}) into
Eq. (\ref{eq: N}), disregarding the weak logarithmic dependence of $\Theta$
and $\Delta e$ on $n$:
\begin{equation}
N = \frac{\Theta\Delta E}{\Delta e} \propto \frac{(\beta\Delta E)
(\Delta E)}{1/\beta} = \beta^2(\Delta E)^2\propto
\mathrm{poly}(n),
\end{equation}
by the assertion that the energy bandwidth is a polynomial
function of the number of spins in our system.
This result shows that in the absence of error in the calculated Fourier
components of the density of states, the free energy per spin can be
determined efficiently to bounded error.

\section{Error Analysis: Fourier Components}

We now consider random errors in the individual values of
$\tilde{g}_\ell$, which may arise from imprecise implementation of
logic gates, or noise in the measurement process.  Treating
$\mathrm{Re}(\tilde{g}_\ell)$ and $\mathrm{Im}(\tilde{g}_\ell)$ as
random variables, these fluctuations are modelled by their
variances.  We assume that the variances $\sigma^2_g$ are
independent \footnote{The assumption is
reasonable if we consider the error for the ensemble quantum
algorithm to arise from noise in the measurement process, as
discussed below.} of $\ell$.  In this section, the dependence
of the maximum allowable value of $\sigma^2_g$ on $n$ such that
Eq. (\ref{Eq:I.3}) is maintained is derived.

As the estimate for the partition function $\tilde{Z}$ is a linear
combination of the independent random variables
$\mathrm{Re}(\tilde{g}_\ell)$ and $\mathrm{Im}(\tilde{g}_\ell)$,
the variance of $\tilde{Z}$ can be calculated from Eq.
(\ref{partition function from f samples}):
\begin{equation}
\sigma_{\tilde{Z}}^2 = 4^{n+1}\left(\frac{\Delta
t}{2\pi\beta}\right)^2(1-e^{-\beta\Delta E})^2
\sigma_g^2\sum_{\ell>0}^{N/2}\frac{b_{\Theta,\ell}^2}{1+(t_\ell/\beta)^2}.
\label{Eq:V.1}
\end{equation}
If we assume that $\tilde{Z}$ is Gaussian-distributed \footnote{
The validity of this assumption is dependent on the nature of the
noise source.  It is exact if the probability distribution
functions for each of the $\mathrm{Re}(\tilde{g_\ell})$ and
$\mathrm{Im}(\tilde{g_\ell})$ are themselves Gaussian.}, then the
probability of $\tilde{Z}$ deviating from its exact value $Z$ can
be related to the variance.  Thus, the sum in Eq. (\ref{Eq:V.1}) is
evaluated by making two simplifications.  First, we model the
windowing function $b_\Theta(t)$ as a Gaussian.  Recall that $b_\Theta(t)$
is constructed by the convolution of $\Theta$ rectangular windows of
width $T_0$.  In the
limit of large $\Theta$, $b_\Theta(t)$ may be approximated
\footnote{As the Fourier transform of $b_\Theta(t)$ is the broadening
function $b_\Theta(E)$, which is of unit area, $b_\Theta(t=0)=1$.} by
\begin{equation}
b_\Theta(t) \approx e^{-t^2/2\nu^2},
\end{equation}
where $\nu^2 = \Theta T_0^2/12$.  Although this approximation
overestimates $b_\Theta(t)$ away from $t=0$, the fractional error in Eq.
(\ref{Eq:V.1}) incurred by the approximation is less than $5\times
10^{-3}$ for $\Theta>40$. Second, it is assumed that $\beta/\Delta
t = \beta\Delta E/2\pi \gg 1$.  The energy bandwidth is thus much
larger than the thermal energy.  This condition assures that the
sum can be well-approximated by the integral
\begin{align}
\sigma_{\tilde{Z}}^2&\approx 4^{n+1}\left(\frac{\Delta
t}{2\pi\beta}\right)^2\sigma_g^2
\int_0^\infty\frac{e^{-t^2/\nu^2}}{1+t^2/\beta^2}
\frac{dt}{\Delta t} \nonumber \\
&=\frac{4^n\sigma_g^2}{\beta\Delta E}
e^{\beta^2/\nu^2}[1-\mathrm{erf}(\beta/\nu)]. \label{eq:V.2}
\end{align}

Eq. (\ref{eq:V.2}) indicates that the standard deviation of
$\tilde{Z}$ scales exponentially \footnote{Although the term
$e^{\beta^2/\nu^2}[1-\mathrm{erf}(\beta/\nu)]$ is a weakly
decreasing function of $n$, the $4^n$ dependence dominates.} with
$n$; \emph{i.e.}, as $2^n$.  Note that the exact partition function
$Z$ will typically be a more slowly increasing function of $n$. If
the energy eigenvalues are limited to the domain $[0,\Delta E]$,
then $2^n$ is an upper bound for the value of the partition
function (achieved at infinite temperature, or if all eigenstates
are degenerate with zero energy).  Consider two simple examples.  For
the case of $n$ non-interacting spins in a magnetic field with
Zeeman energy $h$, $Z=(1+e^{-\beta h})^n<2^n$;  for a linear chain Ising
model in zero magnetic field, described by Eq. (\ref{eq: Ising model}),
$Z=(1+e^{-2\beta J})^n$ for
periodic boundary conditions.  Thus, if the distribution function
for $\tilde{Z}$ is Gaussian, one expects that the standard deviation increases
exponentially faster \footnote{The mean of $\tilde{Z}$ is not strictly $Z$
due to the broadening error, but may be bounded to an arbitrarily
small region about $Z$ by the techniques of the previous
section.} than the mean $Z$.

The above result may be used to derive a condition on $\sigma_g^
2$ such that the error bound on the free energy per spin is fulfilled.
By Eq. (\ref{Eq:I.1}), the condition $|\tilde{F}-F|<\gamma k_B\Theta$ in
Eq. (\ref{Eq:I.3}) is equivalent to
\begin{equation}
Ze^{-\gamma n} < \tilde{Z} < Z e^{\gamma n}.
\end{equation}
Assuming a Gaussian distribution for $\tilde{Z}$ centered about
$Z$,
\begin{align}
\epsilon&=1-\mathrm{Prob}(Ze^{-\gamma n} < \tilde{Z} < Z e^{\gamma n}) \\
&=\frac{1}{2}\left\{
\mathrm{erfc}\left[\frac{Z(e^{\gamma n}-1)}{\sqrt{2}\sigma_{\tilde{Z}}}\right]
+\mathrm{erfc}\left[\frac{Z(1-e^{-\gamma n})}{\sqrt{2}\sigma_{\tilde{Z}}}\right]
\right\}. \nonumber
\end{align}
This result can be simplified if we consider the limit $\gamma n \ll 1$, such
that $e^{\pm\gamma n}\approx 1\pm\gamma n$; \emph{i.e.}, for small desired
absolute error in the free energy relative to the number of spins:
\begin{equation}
\epsilon = \mathrm{erfc}\left(\frac{Z\gamma n}{\sqrt{2}\sigma_{\tilde{Z}}}\right)
\approx \mathrm{erfc}\left(\sqrt{\frac{\beta\Delta E}{2}}
\frac{Z\gamma n}{2^n\sigma_g}\right).
\end{equation}
The argument of the $\mathrm{erfc}(\cdot)$ function must be order unity or
larger for $\epsilon < 0.1$, so
\begin{equation}
\sigma_g^2 = O\left(\frac{Z^2\mathrm{poly}(n)}{4^n}\right).
\label{eq:V.3}
\end{equation}
By the above argument, the variance in the measured Fourier components must
decrease exponentially with $n$.

The condition on $\sigma_g^2$ is likely to translate into an exponentially
scaling computation time for the overall calculation.  For example, consider as
a quantum computer an ensemble of spin-1/2 nuclei, where readout is performed by
measuring the voltage induced in a pickup coil by free induction.  A source of
error in the measured Fourier components is the Johnson-Nyquist voltage noise
due to the resistance of the coil \cite{Hoult1}.  The variance in the observed
voltage -- and thus in the estimates for $\mathrm{Re}(g_\ell)$ and
$\mathrm{Im}(g_\ell)$ -- is inversely proportional to the measurement time.
Thus, Eq. (\ref{eq:V.3}) implies that an exponentially long measurement time is
required to satisfy the condition in Eq. (\ref{Eq:I.3}).

\section{Conclusion}

We examined the applicability of spectral quantum algorithms for
the calculation of the free energy of spin lattice models.
Provided that the time-evolution operator for the system is
decomposable into an efficient number of elementary gates, an
ensemble quantum algorithm exists to generate estimates of the
density of states by calculating individual Fourier components of
$\rho(E)$.  We analyzed the efficiency of this algorithm in
calculating the free energy per spin of the system to bounded
absolute error.

The error in the calculated free energy arises from the
calculation of only a discrete number of Fourier components
$f_\ell$, as well as from deviations in the measured values of
$f_\ell$ due to statistical errors.  The first source of error,
attributable to broadening in the estimated density of states, was
shown to lead to bounded error with a number of Fourier components
that is polynomial in $n$. Thus, if the components $f_\ell$ are
known exactly, the spectral algorithm is an efficient means to
calculate the free energy per spin.  However, the effect of random
deviations in the calculated values of $f_\ell$ grows with
increasing $n$.  As the size of the system increases, the maximum
tolerable variance in measured Fourier components decreases as
$Z^2/4^n$ for large $n$ and small absolute error.  As an upper
bound for the partition function is $2^n$, the spectral algorithms
are not an efficient method to determine $F$ in the presence of
statistical errors in $f_\ell$.

\begin{acknowledgments}
This work is partially supported by the DARPA QuIST program.  CPM
acknowledges the support of the PACCAR Inc. Stanford Graduate Fellowship.
\end{acknowledgments}

\appendix
\section{Proofs of Lemmas 1 and 2}
\emph{Proof of Lemma 1:}  A lower bound is first derived for
\begin{equation}
I \equiv \int_{-\infty}^{\infty} \left[\sinc(x)\right]^\Theta dx =
\int_{-\infty}^{\infty} e^{\Theta\ln\left[\sinc(x)\right]}dx.
\end{equation}
We exclude infinitesimal regions around $x=m\pi$ ($m \in \mathcal{Z}$)
from the integral to avoid divergence of the logarithm; as $\mathrm{sinc}(x)$
approaches a finite value in these regions, the contribution of these
regions to the integral can be made arbitrarily small.

Using a series expansion for $\ln\left[\sinc(x)\right]$
\cite{Jeffrey1},
\begin{align}
\ln\left[\sinc(x)\right] &= -\frac{x^2}{6} - \sum_{k=2}^\infty
\frac{x^{2k}}
{k\pi^{2k}}\left(\sum_{n=1}^\infty\frac{1}{n^{2k}}\right) \nonumber \\
&> -\frac{x^2}{6} - \left(\frac{\pi^2}{6}\right)\sum_{k=2}^\infty
\frac{x^{2k}} {k\pi^{2k}}.
\end{align}
Thus,
\begin{align}
I > \int_{-\infty}^{\infty} e^{-\Theta x^2/6}
e^{-\frac{\Theta\pi^2}{6}\sum_{k=2}^\infty
\frac{x^{2k}}{k\pi^{2k}}}dx.
\end{align}
The integrand is positive over the entire domain of $x$,
and both exponential factors monotonically decrease with
$|x|$.  Thus, one may place a lower bound on $I$ by reducing the
limits of integration to any finite interval, such as
$|x|<\sqrt{6/\Theta}$.  Thus,
\begin{align}
I > e^{-\frac{\Theta\pi^2}{6}\sum_{k=2}^\infty
\frac{(6/\Theta)^{k}}{k\pi^{2k}}}
\int_{-\sqrt{6/\Theta}}^{\sqrt{6/\Theta}}e^{-\Theta x^2/6}dx.
\end{align}
The integral is  $\sqrt{6\pi/\Theta}\ \mathrm{erf}(1)$.  The
summation can be performed explicitly to yield
\begin{align}
I &>
e^{1+\pi^2\Theta\ln(1-6/\pi^2\Theta)/6}\sqrt{\frac{6\pi}{\Theta}}\mathrm{erf}(1)
\nonumber \\
&=
e\left(1-\frac{6}{\Theta\pi^2}\right)^{\Theta\pi^2/6}\mathrm{erf}(1)
\sqrt{\frac{6\pi}{\Theta}}
\nonumber\\&>e\left(1-\frac{6}{\pi^2}\right)^{\pi^2/6}\mathrm{erf}(1)
\sqrt{\frac{6\pi}{\Theta}}.
\end{align}
where we make use of the fact that $(1-1/x)^x$ is a monotonically
increasing function for $x>1$.

This lower bound for $I$ is used to establish an upper bound
for $\alpha_\Theta$.
\begin{align}
\alpha_\Theta &= \frac{1}{\int_{-\infty}^{\infty}\left[\sinc
\left(\frac{\pi E}{\Delta e}\right)\right]^\Theta dE} \nonumber \\
&= \frac{\pi}{\Delta e \ I} \nonumber \\
&< \frac{\pi}{\Delta e} \left(c \sqrt{\frac{\Theta}{6\pi}}\right),
\end{align}
where $c$ is defined as
\begin{equation}
c \equiv \frac{1}{e}\left(\frac{1}{1-6/\pi^2}\right)^{\pi^2/6}
\frac{1}{\mathrm{erf}(1)} \approx 2.0367.
\end{equation}

\emph{Proof of Lemma 2:}
For $\Theta$ even, $b_\Theta(E)$ is a non-negative function with
unit area.  If one treats $b_\Theta(E)$ as a probability density
function, one can use the Markov inequality to bound the area
outside of the main lobe.

Consider a random variable $Y$ with support $y \geq 0$;
\emph{i.e.,} $Y$ only takes non-negative values.  Markov's
inequality bounds the probability of deviations from the mean:
\begin{equation}
\mathrm{Pr}\left(Y\geq\delta\right) \leq \frac{E(Y)}{\delta},
\end{equation}
where $E(Y)$ is the expectation value of $Y$.  Define a second
random variable $X$, such that $Y = [X-E(X)]^m$, where $m$ is an
even integer.  Then,
\begin{align}
\mathrm{Pr}\left\{[X-E(X)]^m \geq \delta\right\} &\leq \frac
{E\left\{[X-E(X)]^m\right\}}{\delta}\nonumber\\
\Rightarrow\mathrm{Pr}\left\{|X-E(X)| \geq \epsilon\right\} &\leq
\frac{E\left\{[X-E(X)]^m\right\}}{\epsilon^m} \label{A: Chebyshev
inequality}
\end{align}
This bound is expressed in terms of the $m^{\mathrm{th}}$ central
moment of $X$, if it exists.  The result reduces to Chebyshev's
inequality for $m=2$.

Note that if one treats $b_\Theta(E)$ as a probability distribution
function for a zero-mean random variable $E$, the above inequality
provides a bound for the area outside the main lobe (\emph{i.e.,}
$\epsilon = \Delta e$).  The central moment is evaluated for $m =
\Theta-2$.
\begin{align}
E\left\{[X-E(X)]^m\right\} &= \int_{-\infty}^{\infty} E^{\Theta-2}
b_\Theta(E)dE \nonumber \\
&= \alpha_\Theta\left(\frac{\Delta e}{\pi}\right)^{\Theta-1}
\int_{-\infty}^{\infty} \frac{\sin^\Theta x}{x^2}dx \nonumber \\
&\leq \alpha_\Theta\left(\frac{\Delta e}{\pi}\right)^{\Theta-1}
\int_{-\infty}^{\infty} \frac{\sin^2 x}{x^2}dx \nonumber \\
&= \alpha_\Theta\left(\frac{\Delta e}{\pi}\right)^{\Theta-1}\pi
\end{align}
If we define the area outside the main lobe as
\begin{equation}
A_{\text{side}} \equiv 1-\int_{-\Delta e}^{\Delta e}b_\Theta(E)dE,
\end{equation}
then,
\begin{equation}
A_{\text{side}} = \mathrm{Pr}\left\{|X-E(X)| \geq \Delta e\right\} \leq
\frac{\alpha_\Theta\Delta e}{\pi^{\Theta-2}}. \
\label{A: lemma 2 almost done}
\end{equation}
Combining Eq. [\ref{A: lemma 2 almost done}] with Lemma 1, we find
\begin{equation}
A_{\text{side}} < \frac{c}{\pi^{\Theta-3}} \sqrt{\frac{\Theta}{6\pi}}.
\end{equation}

\end{document}